\documentclass[prl,aps,amsfonts,amssymb,draft,floats,twocolumn]{revtex4}
\usepackage{epsf}

\newcommand{\be}{\begin{equation}}
\newcommand{\ee}{\end{equation}}

\begin{document}

\title{Comment on ``Vortices with Fractional Flux in Two-Gap Superconductors
and Extended Faddeev Model'', 
``Phase Diagram of Planar  $U(1)\times U(1)$ Superconductor''
and ``Flux Noise in ${\rm MgB}_2$ Thin Films''
}

\author{D.~A.~Gorokhov }

\affiliation{{Laboratory~of~Atomic and Solid State Physics, Cornell University, Ithaca, NY 14853-2501, USA}\\
{ {\rm September 21, 2003}  }
}

\vskip-0.7cm
\begin{abstract}
I show that recent theoretical~\cite{Babaev,Babaev1} and experimental~\cite{Festin}
claims about the possibility of the Berezinskii-Kosterlitz-Thouless
(BKT) transition of ``fractional vortices'' in thin films of ${\rm MgB}_2$ 
are inconsistent with the parameters of the electron-phonon interaction  in ${\rm MgB}_2$.
\end{abstract}
\maketitle

\vskip0.5cm{
{\bf 1.~Theory.}
In Ref.~\cite{Babaev} Babaev investigates vortices in
two-gap superconductors (TGS). He finds flux lines carrying a fraction
of the flux quantum $\Phi_0$ and suggests an experiment for the
observation of the BKT-transition of such topological excitations in
thin films of TGS.  It is claimed that ``the results could be relevant
for the newly discovered two-gap superconductor ${\rm MgB}_{2}$''.  I
show that, {\it i}) in~\cite{Babaev} the free energy is treated
improperly and the vortex solutions obtained are invalid for ${\rm
MgB}_{2}$; {\it ii}) the careful analysis excludes the possibility to
observe these vortices in the suggested in \cite{Babaev} experiment on
${\rm MgB}_{2}$.
}

The free energy density $F$ for a thin film made of a TGS with a
nonzero Josephson coupling (JC) $g$ (for the microscopic Hamiltonian,
see Eq.~(1) of \cite{Suhl}) can be written in the form $F= F_1 + F_2 +
(g/2) \left (\Delta_1\Delta_2^{*} + \Delta_1^{*}\Delta_2\right )$,
where $\Delta_i = |\Delta_i | e^{i\phi_i}$, $i=1,2$ are the
superconducting order parameters of the two condensates and \be F_i =
-\frac{\alpha_i}{2}{\left |\Delta_i\right |}^2 +
\frac{\beta_i}{4}{\left |\Delta_i\right |}^4 +
\frac{\gamma_i^{kl}}{2}\Pi_k\Delta_i\Pi_l^{*}\Delta_i^{*},
\label{free_energy}
\ee with ${\bf \Pi} = \nabla + 2\pi i {\bf A}/\Phi_0$ and $k,l=x,y,z$.
Remarkably, the above free energy $F$ describes equally well a system
of two superconducting layers situated close to each other.  The
topological excitations in layered superconductors such as Bi- and
Tl-based high-$T_c$ superconductors consist of pancake
vortices~\cite{Clem} describing the singular phase fluctuations in
each layer. In a zero magnetic field and for $g=0$, pancake vortices
in the same layer interact logarithmically at all distances $R$
exceeding the core size and exhibit a BKT-like phase
transition~\cite{Clem}. Also,
individual pancake vortices carry a fraction of the magnetic flux
quantum~\cite{Clem}.  Hence, in a thin film of a two-gap superconductor,
the elementary topological excitations found  in
Ref.~\cite{Babaev} are equivalent to pancake vortices in a two-layer
system. E.g., the vortex ``$\Delta \phi_1 =2\pi , \Delta
\phi_2 = 0$'' defined in Ref.~\cite{Babaev} corresponds to one pancake
in the first layer (condensate) and to no pancakes in the second one.

For $g\ne 0$ there is a new lengthscale $\Lambda$ 
such that for $R > \Lambda$ vortex-antivortex pairs are attracted with
a potential {\it linear} in $R$ and thus exhibit confinement, i.e.,
the BKT-transition is quenched. However, if $\Lambda$ is large,
a BKT-like
crossover smeared on the scale $\Lambda$ can be observed
\cite{Artemenko}.

{\it i)} It is claimed in \cite{Babaev} that if $L<\Lambda$ ($L$ is the sample
size and $\Lambda$
is defined as ``inverse mass of the field $n_1$'' in Ref.~\cite{Babaev}), 
JC can be neglected. Solutions for fractional vortices
have been obtained on all lengthscales (even exceeding $\lambda$, the
magnetic field penetration length).
In fact, the lengths $\Lambda$, $\xi_i$'s and $\lambda$ are
characteristic for every superconductor and cannot be chosen
arbitrarily.  For ${\rm MgB}_2$ the condition
$\Lambda < {\xi_1(T)}, {\xi_2(T)}$ holds 
($\xi_i$'s are the coherence lengths of the condensates),
i.e. even a single vortex
cannot fit inside a superconductor of size $L < \Lambda$ contrary
to what is described in
~\cite{Babaev}.

{\it ii}) Let us then estimate the parameters $\Lambda$ and $\xi_i$
for the TGS and show that for ${\rm MgB}_2$ the length $\Lambda$ is
not large enough to neglect JC and observe the BKT-transition.  Let
the film be parallel to the $xy$-plane. We assume
$\gamma_i^{xx}=\gamma_i^{yy}\equiv\gamma_i$ and $\gamma_i^{kl}
\propto\delta_{kl}$. These assumptions are valid for ${\rm
MgB}_2$. The indices $i=1,2$ correspond to the $\sigma$- and
$\pi$-bands respectively.  Neglecting spacial fluctuations of
$|\Delta_i |$ (thin film) and, varying the free energy $F$ with
respect to $\phi_1$ and $\phi_2$, we obtain the equation
$\Lambda^2\Delta\phi = - \sin\phi$, with $\Delta = \partial_x^2 +
\partial_y^2$, $\phi = \phi_1 - \phi_2$, and \be \Lambda^2 =
\frac{2}{g}\frac{1}{|\Delta_1| |\Delta_2|} \frac{\gamma_1 |\Delta_1|^2
\gamma_2 |\Delta_2|^2} {\gamma_1 |\Delta_1|^2 + \gamma_2
|\Delta_2|^2}.
\label{J_coupling}
\ee Solving this equation for a vortex-antivortex pair and
substituting the solution back into $F$ one can find that for
$R>\Lambda$ the interaction is linear in $R$.  The coupling $g$ is
determined through $g = \lambda_{12} N_1/ {\left
(\lambda_{11}\lambda_{22} - \lambda_{12}\lambda_{21}\right )}$, with
$\lambda_{11}\approx 0.81$, $\lambda_{22}\approx 0.285$,
$\lambda_{12}\approx 0.119$, $\lambda_{21}\approx 0.09$,
$N_2/N_1\approx 1.3$, see \cite{Gurevich,Mazin}.  The in-plane
coherence lengths $\xi_i (T)$ can be estimated in the dirty limit as
$\xi_i^2 (T)\simeq \gamma_i/\beta_i\Delta_i^2$, with $\beta_i =
7\zeta(3)N_i/8\pi^2 T_c^2$~\cite{Gor'kov} and $N_i$ the density of
states.  Near critical temperature $T_c$, $\Delta_1/\Delta_2\approx
6$~\cite{Gurevich}.  Using $a_1a_2/(a_1 + a_2) < {\rm min}\{ a_1 ,
a_2\}$, with $a_i = \gamma_i {|\Delta_i |}^2$ and substituting the
above values for the parameters into (\ref{J_coupling}) we find that
$\Lambda < 1.55\xi_i ( T_{\rm BKT}) \left [\Delta_i (T_{\rm
BKT})/T_c\right ]$.  Since $T_c - T_{\rm BKT}\ll T_c$, then $\Delta_i
(T_{\rm BKT})\ll T_c$ and the lengthscale $\Lambda$ is not large
enough.  The same conclusion appears for the pure
limit\cite{Geilikman,Zhitomirsky}.

In conclusion, the BKT-physics described in~\cite{Babaev} is
well-established for layered superconductors
\cite{Clem,Artemenko} but is not applicable to the TGS ${\rm MgB}_2$.
The results of paragraph {\it ii} above also apply to Ref.~\cite{Babaev1} where
the BKT-transition of ``fractional vortices'' is discussed and it is claimed
that ``it is very likely that in ${\rm MgB}_2$ the JC is small''.
 In fact,
the JC strength $g$ should be smaller at least by a factor of $10^3$
in order to observe the BKT-like crossover in ${\rm MgB}_2$.

{\bf 2.~Experiment.}
In a recent paper~\cite{Festin} Festin {\it et al.} investigate
flux noise in ${\rm MgB}_2$ films.  They observe sharp
BKT-transitions and interpret the results ``in terms of  vortices carrying an
arbitrary fraction of a flux quantum'' discussed in Refs.~\cite{Babaev,Babaev1}. 
Festin {\it et al.}  claim that their result give ``support for the presence of 
fractional vortices and because of this the existence of a BKT transition
in comparably thick ${\rm MgB}_2$ films''. 
The authors exclude the possibility of the
BKT-transition of conventional vortices carrying the flux $\Phi_0$.
I show that: {\it i)} The BKT-transition of ``fractional''
(pancake) vortices cannot be observed in ${\rm MgB}_2$ films; {\it
ii)} The possibility of the BKT-transition of ordinary vortices cannot
be excluded from the experimental data presented in~\cite{Festin}.

{\it i)} This statement follows from the theoretical analysis
presented above, see Section 1.

{\it ii)} It is claimed in \cite{Festin} that ``for ordinary Abrikosov
vortices, the BKT transition would not
be observed experimentally since there is an exponential cut-off in the logarithmic
interaction for vortex separations being larger than the effective penetration depth''.
This statement is incorrect.  
In fact, in this situation
one can observe a sharp BKT-like crossover if $\lambda^2(T)/d \gg
\xi(T)$, with $d$ the film thickness and  
$\xi(T)$ the core size of the vortex carrying the flux $\Phi_0$,
as
vortices have enough room to explore the logarithmic interaction.
 In the whole temperature range (except for
temperatures $T$ very close to $T_{\rm BKT}$) a superconducting
sample exhibits critical behavior indistinguishable from the 
BKT-transition. Only when the BKT-correlation length 
$\xi_{\rm BKT}(T)$ becomes of the order of $\lambda^2(T)/d$, the transition
becomes smeared. For the samples used in the experiment~\cite{Festin}
the condition 
$\lambda^2(T_{\rm BKT})/d >> \xi (T_{\rm BKT})$ is clearly
satisfied since $\lambda (T_{\rm BKT}) \gg  \xi (T_{\rm BKT})$
and  $\lambda (T_{\rm BKT}) \gg d$. This indicates that the possibility
of the BKT-transition (sharp crossover) of conventional vortices
cannot be excluded.

In conclusion, the experimental results~\cite{Festin}
cannot be interpreted in terms
of the BKT-transition of ``fractional'' (pancake) vortices
and the possibility of the BKT-transition of conventional
vortices carrying the flux $\Phi_0$ cannot be excluded based on the 
data presented in~\cite{Festin}.

The present work was supported by the Packard foundation.

\end{document}